\documentstyle[12pt]{article}
\begin{document}

\begin{flushright}
UT-Komaba 99-6  \\
hep-lat/9905001
\end{flushright}

\begin{center} 
{\Large{\bf  Pion Mass and the PCAC Relation in the Overlap
Fermion Formalism: }}\\
\vskip 0.15cm
{\Large{\bf Gauged Gross-Neveu Model on a Lattice }}
\vskip 1.5cm

{\Large  Ikuo Ichinose\footnote{e-mail 
 address: ikuo@hep1.c.u-tokyo.ac.jp}and 
 Keiichi Nagao\footnote{e-mail
 address: nagao@hep1.c.u-tokyo.ac.jp}}  
\vskip 0.5cm
 
Institute of Physics, University of Tokyo, Komaba,
  Tokyo, 153-8902 Japan  
 
\end{center}

\vskip 4cm
\begin{center} 
\begin{bf}
Abstract
\end{bf}
\end{center}
We investigate chiral properties of the overlap lattice
fermion by using solvable model in two dimensions,
the gauged Gross-Neveu model.
In this model, the chiral symmetry is spontaneously broken
in the presence of small but finite fermion mass.
We calculate the quasi-Nambu-Goldstone(NG) boson mass 
as a function of the bare fermion mass
and two parameters in the overlap formula.
We find that the quasi-NG boson
mass has desired properties as a result of 
the extended chiral symmetry found by L\"uscher.
We also examine the PCAC relation and find that it is satisfied
in the continuum limit.
Comparison between the overlap and Wilson lattice fermions  
is made.

\newpage
\setcounter{footnote}{0}

Species doubling is a long standing problem in the lattice fermion
formulation.
Wilson fermion is the most suitable formulation\cite{Wilsonterm} and it is used
in most of the numerical studies of lattice gauge theory.
However in order to reach the desired continuum limit, fine tunning
must be done with respect to the ``bare fermion mass" and the 
Wilson parameter.

Recently a very promising formulation of lattice fermion
named overlap fermion was proposed by Narayanan and Neuberger\cite{overlap}.
In that formula the Ginsparg and Wilson(GW)
 relation\cite{GinspargWilson} plays a very important
role, and because of that there exists an ``extended" (infinitesimal)
chiral symmetry.

In this paper we shall study or test the overlap fermion by
using the gauged Gross-Neveu model in two dimensions.
This is a solvable model which has similar
chiral properties with QCD$_4$, i.e., chiral symmetry is spontaneously
broken with a small but finite bare fermion mass and pion appears
as quasi-Nambu-Goldstone boson\footnote{Overlap fermion with
a finite fermion mass was recently studied in Ref.\cite{mass}}.
Actually a closely related model was studied on a lattice in order to test
properties of the Wilson fermion in the continuum limit\cite{Ichinose}.
Therefore advantage of the overlap fermion becomes clear
by the investigation in this paper.

The model is defined by the following action on the two-dimensional
square lattice with the lattice spacing $a$,
\begin{eqnarray}
S&=& {N \over 2} \sum_{pl} \prod U_\mu(n) 
+ a^2\sum_{n,m}\bar{\psi}(m)D(m,n)\psi(n)
+a^2M_B\sum_n\bar{\psi}\psi(n)   \nonumber \\
&& 
-{a^2 \over \sqrt{N}}\sum_n\Big[\phi^i(n)(\bar{\psi}\tau^i\psi)(n)
+\phi^i_5(n)(\bar{\psi}\tau^i\gamma_5\psi)(n)\Big] \nonumber  \\
&& +{a^2 \over 2g_v}\sum_n\Big[\phi^i(n)\phi^i(n)+
\phi^i_5(n)\phi^i_5(n)\Big],
\label{action1}
\end{eqnarray}
where $U_\mu(n)$ is U(1) gauge field defined on links,
$\psi^l_{\alpha}\; (\alpha=1,...,N,
l=1,...,L)$ are fermion fields with flavour index $l$,
and the matrix $\tau^i\; (i=0,...,L^2-1)$ acting on the flavour
index is normalized as
\begin{equation}
{\rm{Tr}}(\tau^i\tau^k)=\delta_{ik}
\end{equation}
and 
\begin{equation}
\tau^0={1 \over \sqrt{L}}, \; \; \{\tau^i,\tau^j\}=d^{ijk}\tau^k,
\end{equation}
where $d^{ijk}$'s are the structure constants of $SU(L)$.
Fields $\phi^i$ and $\phi^i_5$ are scalar and pseudo-scalar
bosons,respectively.
The covariant derivative in Eq.(\ref{action1}) is defined by the
overlap formula
\begin{eqnarray}
D&=&{1\over a}\Big(1+X{1 \over \sqrt{X^{\dagger}X}}\Big),  \nonumber  \\
X_{nm}&=&\gamma_{\mu}C_{\mu}(n,m)+B(n,m),  \nonumber   \\
C_{\mu}&=&{1 \over 2a}\Big[\delta_{m+\mu,n}U_{\mu}(m)-
\delta_{m,n+\mu}U^{\dagger}_{\mu}(n)\Big],  \nonumber  \\
B(n,m)&=&-{M_0\over a}+{r\over 2a}\sum_{\mu}\Big[2\delta_{n,m}
-\delta_{m+\mu,n}U_{\mu}(m)-\delta_{m,n+\mu}U^{\dagger}_{\mu}(n)\Big],
\label{covD}
\end{eqnarray} 
where $r$ and $M_0$ are 
dimensionless nonvanishing free parameters of the overlap lattice fermion 
formalism\cite{overlap,Neuberger}.
The overlap Dirac operator $D$ does not have the ordinary chiral 
invariance but satisfies the GW relation instead,
\begin{equation}
D\gamma_5+\gamma_5D=aD\gamma_5D.
\label{GWr}
\end{equation}
From (\ref{action1}) it is obvious that the systematic $1/N$ expansion
is possible and we shall employ it.

The action (\ref{action1}) contains the bare fermion mass $M_B$
which explicitly breaks the chiral symmetry.
This bare mass also breaks the following infinitesimal 
transformation, which was discovered by L\"uscher\cite{Luscher} and we call
``extended chiral symmetry",
\begin{eqnarray}
&& \psi(n) \rightarrow \psi(n)+\tau^k\theta^k\gamma_5\Big\{
\delta_{nm}-{1 \over 2}aD(n,m)\Big\}\psi(m),  \nonumber  \\
&& \bar{\psi}(n) \rightarrow \bar{\psi}(n)+\bar{\psi}(m)
\tau^k\theta^k\gamma_5\Big\{\delta_{nm}-{1 \over 2}aD(n,m)\Big\}
  \nonumber  \\
&&\phi^i(n) \rightarrow \phi^i(n)+d^{ikj}\theta^k\phi^j_5(n),  \nonumber  \\
&&\phi^i_5(n) \rightarrow \phi^i_5(n)-d^{ikj}\theta^k\phi^j(n),
\label{extended}
\end{eqnarray}
where $\theta^i$ is an infinitesimal transformation parameter.
The above symmetry (\ref{extended}) coincides with the ordinary
chiral symmetry up to $O(a)$.

From the action (\ref{action1}), it is obvious that 
$\phi^i$ and $\phi^i_5$ are composite fields
of the fermions,
\begin{equation}
\phi^i={g_v \over \sqrt{N}}\bar{\psi}\tau^i\psi, \;\;
\phi^i_5={g_v \over \sqrt{N}}\bar{\psi}\gamma_5\tau^i\psi.
\label{phi}
\end{equation}
As in the continuum model, we expect that the field
$\phi^0$ acquires a nonvanishing vacuum expectation value(VEV),
\begin{equation}
\langle\phi^0\rangle=\sqrt{NL}M_s,
\label{VEV}
\end{equation}
and we define subtracted fields,
\begin{eqnarray}
&&\varphi^0=\phi^0-\sqrt{NL}M_s,  \nonumber  \\
&&\varphi^i=\phi^i\; \; (i\neq 0),  \;\; \varphi^i_5=\phi^i_5.
\end{eqnarray}
In terms of the above fields, 
\begin{eqnarray}
S&=& {N \over 2}\sum_{pl}\prod U_\mu(n)+
 a^2\sum_{n,m}\bar{\psi}(m)D_M(m,n)\psi(n) \nonumber \\
&& -{a^2 \over \sqrt{N}}\sum_n\Big[\varphi^i(n)(\bar{\psi}\tau^i\psi)(n)
+\varphi^i_5(n)(\bar{\psi}\tau^i\gamma_5\psi)(n)\Big] \nonumber  \\
&& +{a^2 \over 2g_v}\sum_n\Big[\varphi^i(n)\varphi^i(n)+
2\sqrt{NL}M_s\varphi^{0}(n)+\varphi^i_5(n)\varphi^i_5(n)\Big],
\label{action2}
\end{eqnarray}
where 
\begin{equation}
D_M=D-M, \;\; M=M_B+M_s.
\label{M}
\end{equation}
Obviously, $M$ is the dymanical fermion mass.

From the chiral symmetry and (\ref{VEV}), we can expect that 
quasi-Nambu-Goldstone(NG) bosons appear as a result of the spontaneous
breaking of the chiral symmetry.
They are nothing but $\varphi^i_5$.
The VEV $M_s$ is  determined by the tadpole cancellation
condition of $\varphi^0$.
In order to perform an explicit calculation of the $1/N$-expansion,
it is useful to employ the momentum representation, and also
we introduce the gauge potential $\lambda_{\mu}(n)$ in the usual way,
i.e., $U(n,\mu)=\exp ({ia\over \sqrt{N}}\lambda_{\mu}(n))$.
By using weak-coupling expansion by Kikukawa 
and Yamada\cite{weakexpansion},  
\begin{eqnarray}
D_{nm}&=&\int_p\int_qe^{-ia(qn-pm)}D(p,q),    \\
D(p,q)&=&D_0(p)(2\pi)^2\delta(p-q)+{1 \over a}V(p,q),
\label{overlapF}
\end{eqnarray} 
where $\int_p=\int^{\pi/a}_{-\pi/a}{d^2p\over (2\pi)^2}$
and 
\begin{eqnarray}
D_0(p)&=& {b(p)+\omega(p) \over a\omega(p)}
+{\gamma_\mu i \sin ap_\mu \over a^2\omega(p)},  \label{D0}  \\
V(p,q) &=& 
 \Bigl\{ 
\frac{1}{\omega(p)+\omega(q)} 
\Bigr\}
\Bigl[
X_1(p,q)
-
\frac{X_0(p)}{\omega(p)} X^\dagger_1(p,q) 
\frac{X_0(q)}{\omega(q)}
\Bigr] +...
\end{eqnarray}
\begin{eqnarray}
X_0(p)&=& 
\frac{i}{a} \gamma_\mu \sin a p_\mu 
+ \frac{r}{a} \sum_\mu \left(1-\cos a p_\mu \right) 
-\frac{1}{a} M_0 ,\\
&& \nonumber\\
X_1(q,p)&=& \int_k  (2\pi)^4 \delta(q-p-k)
\, \frac{1}{\sqrt N} \lambda_\mu(k) \, V_{1 \mu}\left(p+\frac{k}{2}\right) , 
\end{eqnarray}
\begin{eqnarray}
a\omega(p)&=& \sqrt{\sin^2(ap_\mu)
+\Big(r\sum_\mu(1-\cos(ap_\mu))-M_0\Big)^2},  \nonumber   \\
ab(p) &=& r\sum_\mu(1-\cos (ap_\mu))-M_0.
\end{eqnarray}
The vertex function is explicitly given as 
\begin{eqnarray}
V_{1 \mu}\left(p+\frac{k}{2}\right) 
&=& i \gamma_\mu \cos a \left(p_\mu+\frac{k_\mu}{2}\right) 
+ r \sin a \left(p_\mu+\frac{k_\mu}{2}\right) \nonumber \\
&=& \frac{\partial}{\partial p_\mu} X_0 \left(p+\frac{k}{2}\right).
\end{eqnarray}

From (\ref{D0}), the tree level propagator is obtained as 
\begin{eqnarray}
D^{-1}_{M(0)}&=& {a\{b(p)+(1-Ma)\omega(p)\} -i\gamma_{\mu}
\sin (ap_\mu) \over \omega(p)\{1+(1-Ma)^2\}+2b(p)(1-Ma)\}}  \nonumber  \\
&\equiv& {A_{\mu}(p)\gamma_\mu+B(p) \over J(p)}.
\label{propa}
\end{eqnarray}
From (\ref{action2}) and (\ref{overlapF}),
\begin{eqnarray}
{M_s \over g_v}&=&-\int_k
{\rm{Tr}}\Big( D^{-1}_{M(0)}(k)\Big) \nonumber  \\
&=&-\int_k {2a\{b(k)+
(1-Ma)\omega(k)\} \over \omega(k)\{1+(1-Ma)^2\}+2b(k)(1-Ma)}.
\label{Ms}
\end{eqnarray} 

Effective action of $\varphi^i$, $\varphi^i_5$ and the gauge field
$\lambda_\mu(n)$ is obtained by integrating out the 
fermions,
\begin{equation}
  e^{-S_{eff}}=\int [D\bar{\psi}D\psi] e^{-S}.
\end{equation}
Especially we are interested in $\varphi_5$ and $\lambda_\mu$
part of the effective action, because $\varphi_5$ is the 
quasi-NG boson(pion) and its coupling with the gauge boson is
related with anomaly.
We define 
\begin{equation}
S^{(2)}_{eff}[\varphi_5]=\int_k{1 \over 2}
\varphi^i_5(-k)\Gamma^5_{ij}(k^2)\varphi^j_5(k)
\label{Svphi}
\end{equation}
where
\begin{eqnarray}
\Gamma^5_{ij}(k^2)&=&\delta_{ij}\Big[{1 \over g_v}+
\int_k {\rm{Tr}}[\gamma_5\langle \psi(k-p)\bar{\psi}(k-p)\rangle
\gamma_5\langle \psi(k)\bar{\psi}(k)\rangle]\Big]  \nonumber  \\
&=&\delta_{ij}\Big[\epsilon+2k^2M_0^2 A(k^2;M)\Big].
\label{Gamma5}
\end{eqnarray}
Parameter $\epsilon$ in $\Gamma^5_{ij}$ (\ref{Gamma5}) is proportional
to the pion mass and measures the derivation from the limit of the exact
chiral symmetry.

In the leading order of the $1/N$, 
\begin{eqnarray}
\epsilon &=& -{2\over a^2} \int^\pi_k {1 \over M_s[\omega(k/a)\{
1+(1-Ma)^2\}+2b(k/a)(1-Ma)]^2}  \nonumber  \\
&& \times \Big[a\{b(k/a)+(1-Ma)\omega(k/a)\}\{\omega(k/a)(1+(1-Ma)^2)
+2b(k/a)(1-Ma)\}  \nonumber  \\
&& \; +M_s\{\sin^2k_\mu+a^2(b(k/a)+(1-Ma)\omega(k/a))^2\}\Big]
\nonumber   \\
&=&{M_BM_0^2  \over M_s}\Big[ -\ln (M_0M^2a)+\mbox{const.}\Big]
+O(a),
\label{epsilon}
\end{eqnarray}
where $\int^\pi_k=\int^\pi_{-\pi}{d^2k \over (2\pi)^2}$ and 
we took the continuum limit to obtain the last line of
(\ref{epsilon}).
From (\ref{epsilon}), $\epsilon \propto M_B+O(a)$ and 
therefore the limit $M_B \rightarrow 0$ is considered as the chiral
limit.
It is instructive to compare the above result with that in the 
continuum theory and the lattice model with the Wilson fermions.
The corresponding expression of $\epsilon$ in the continuum theory
is given as
\begin{equation}
\Big[\epsilon\Big]_{cont}={2M_B \over M_s}{1 \over 4\pi}
\ln \Big({\Lambda^2\over M^2}\Big)+O(1/\Lambda),
\label{cont}
\end{equation}
where $\Lambda$ is the momentum cutoff.
It is obvious that $\epsilon$ in the overlap formalism 
has a very close resemblance to that in the continuum theory.
On the other hand, the corresponding expression 
in the Wilson fermion formalism was obtained
in Ref.\cite{Ichinose} as follows,
\begin{eqnarray}
\Big[\epsilon\Big]_W&=&-{4r_W \over M_sa}L(r_W)
+{2M_B\over M_s}\int^\pi_k{1\over I(k)},  \nonumber  \\
I(k)&=& \sum_\mu \sin^2k_\mu +\Big( -2r_W\sum\sin^2{k_\mu \over 2}
+Ma\Big)^2,  \nonumber   \\
L(r_W)&=&\int^\pi_k {\sum\sin^2(k_\mu/2) \over I(k)},
\label{wilson}
\end{eqnarray}
where $r_W$ is the Wilson parameter.   
Therefore it is obvious that the fine tuning of the ``bare mass" $M_B$
and the Wilson parameter $r_W$ is required in order to reach
the chiral limit.
In this sense, the overlap fermion is better than the Wilson fermion. 

It is also straightforward to calculate $A(k^2;M)$ in (\ref{Gamma5}),
\begin{eqnarray}
A(k^2;M)&=& {1 \over 4\pi\sqrt{k^2(k^2+\mu^2)}}
\ln {k^2+2\mu^2+\sqrt{(k^2+2\mu^2)^2-4\mu^4} \over 
k^2+2\mu^2-\sqrt{(k^2+2\mu^2)^2-4\mu^4}}   \nonumber  \\
&\rightarrow& {1 \over 4\pi\mu^2}+(k^2).
\label{Ak}
\end{eqnarray}
where $\mu=M_0M$ 
and therefore the pion mass is given as $m^2_\pi=2\pi M^2\epsilon$.

There exists a mixing term of the gauge boson $\lambda_\mu$ and 
the pion $\varphi^0_5$,
\begin{equation}
S^{(2)}_{eff}[\lambda_\mu,\varphi^0_5] =-2\sqrt{L}
M^2_0M\int_k\sum \lambda_\mu(-k)
\epsilon_{\mu\nu}k_\nu A(k^2;M)\varphi^0_5(k),
\label{mixing}
\end{equation}
which is identical with the continuum calculation.
This mixing term is related to the discussion of the  U(1) problem 
in QCD$_4$ and the above result suggests that the correct anomaly 
appears in the Ward-Takahashi identity of the axial-vector current.

We shall examine the PCAC relation.
By changing variables
as follows in the path-integral representation of 
the partition function\footnote{We employ this form of change
of variables instead of that is given by (\ref{extended}).
This is merely for technical reason here.},
\begin{eqnarray}
&& \psi(n) \rightarrow \psi(n)+\tau^k\theta^k(n)\gamma_5\Big\{
\delta_{nm}-aD(n,m)\Big\}\psi(m),  \nonumber  \\
&& \bar{\psi}(n) \rightarrow \bar{\psi}(n)\left\{1+\tau^k\theta^k(n)\gamma_5
\right\},  \nonumber  \\
&&\phi^i(n) \rightarrow \phi^i(n)+d^{ikj}\theta^k\phi^j_5(n),  \nonumber  \\
&&\phi^i_5(n) \rightarrow \phi^i_5(n)-d^{ikj}\theta^k\phi^j(n),
\label{extended2}
\end{eqnarray}
we obtain the Ward-Takahashi(WT) identity,
\begin{eqnarray}
&&\langle \partial_\mu j^k_{5,\mu}(n)-2M(\bar{\psi}\tau^k\gamma_5\psi)(n)
+{2\sqrt{N}\over g_v}M_s\varphi^k_5(n)  \nonumber  \\
&& \; \; \; \;  +D^k_A(n)-\delta^{k0}N\sqrt{L}a
{\rm{Tr}}[\gamma_5D(n,n)]\rangle=0,
\label{WT}
\end{eqnarray}
where the last term comes from the measure of the path integral,
and the explicit form of the current operator $j^k_{5,\mu}$ is 
obtained by Kikukawa and Yamada\cite{axialvector} as follows,
\begin{equation}
j^k_{5,\mu}(n)=\tau^k\sum_{lm}\bar{\psi}(l)K_{n\mu}^5(l,m)\psi(m), \nonumber 
\end{equation}
\begin{eqnarray}
K^5_{n\mu}(l,m)&=&\left\{ K_{n\mu}\frac{H}{\sqrt{H^2}}\right\}(l,m),\\ 
H&=&-\gamma_5 X, 
\end{eqnarray}
\begin{equation}
a K_{n\mu}(l,m)
= \gamma_5 \left\{
\int_{-\infty}^{\infty} \frac{dt}{\pi}
\frac{1}{ (t^2+H^2)}
\left( t^2 W_{n\mu} - H W_{n\mu} H \right)
\frac{1}{ (t^2+H^2) } 
\right\}_{lm}, 
\end{equation}
\begin{equation}
W_{n\mu}(l,m)
= \gamma_5 \left\{ 
\frac{1}{2} \left(\gamma_\mu-1\right)
                 \delta_{nl}\delta_{n+\hat\mu, m} \, U_{n\mu}
+ \frac{1}{2} \left(\gamma_\mu+1\right)
 \delta_{l,n+\hat\mu}\delta_{nm} \, U_{n+\hat\mu,\mu}^\dagger 
\right\}. \nonumber\\
\end{equation}
and the operator $D^k_A(n)$ is given by
\begin{eqnarray}
&& D^k_A(n)=aM\sum_m\bar{\psi}(n)\tau^k\gamma_5D(n,m)\psi(m)  \nonumber  \\
&& \;\;\;
+{a \over \sqrt{N}}\bar{\psi}(n)\Big(\varphi^i(n)\gamma_5-\varphi^i_5(n)\Big)
\tau^i\tau^k\sum_mD(n,m)\psi(m).
\label{jD}
\end{eqnarray}

By integrating out the fermions, the above WT identity is expressed 
in terms of the pions and the gauge field.
Matrix element $D^A_5=\langle \varphi^k_5|D^k_A|0\rangle$ has 
the contribution
from the following two terms,
\begin{eqnarray}
D^A_5(p) &=& \{D^A_5(p)\}^a+\{D^A_5(p)\}^b,   \nonumber\\
\{D^A_5(p)\}^a &=& \sqrt{N}a\int_q
{\rm{Tr}}[\langle \psi(q)\bar{\psi}(q)
\rangle D_0(q)] \times \varphi^k_5(p)  \nonumber  \\
&=&\sqrt{N}a\varphi^k_5(p)\int_q{2(2-Ma)(\omega(q)+b(q))
\over J(q)},  \nonumber  \\
\{D^A_5(p)\}^b &=& -\sqrt{N}aM\int_q\varphi^k_5(p){\rm{Tr}}[\langle
\psi(p+q)\bar{\psi}(p+q)\rangle\gamma_5 \langle
\psi(q)\bar{\psi}(q)\rangle\gamma_5D_0(p+q)]  \nonumber  \\
&=& -\sqrt{N}aM\varphi^k_5(p) \int_q{2 \over J(q)J(q-p)}\Big[
-M\sin aq_\mu \sin a(q-p)_\mu  \nonumber  \\
&& \;\;\; -(2-Ma)(\omega(q)+b(q))a\{b(q-p)+(1-Ma)\omega(q-p)\}\Big].
\label{DA5}
\end{eqnarray}
From (\ref{epsilon}) and (\ref{DA5}), we obtain 
\begin{equation}
D^A_5(p)=-2\sqrt{N}M_s\epsilon \varphi^k_5(p).
\label{DA}
\end{equation}
In a similay way, matrix element $D^A_\mu=\langle \lambda_\mu|D^0_A
|0\rangle$ is evaluated as, 
\begin{eqnarray}
D^A_\mu &=& \{D^A_\mu(p)\}^a+\{D^A_\mu(p)\}^b, \nonumber  \\
\{D^A_\mu(p)\}^a&=&NM\sqrt{L}\int_{qq'}
{\rm{Tr}}[\langle\psi(q)\bar{\psi}(q')
\rangle\gamma_5D_0(p+q')\langle \psi(p+q')\bar{\psi}(p+q')\rangle 
\nonumber  \\
&& \; \; \; \times V(p+q',q)]\delta^{k0}  \nonumber  \\
\{D^A_\mu(p)\}^b&=&-NM\sqrt{L}\int_q{\rm{Tr}}[\langle\psi(q)
\bar{\psi}(q)\rangle\gamma_5\delta^{k0}
V(p+q,q)].
\end{eqnarray}
It is not so difficult to show that the above two terms
cancel with each other and  $D^A_\mu(p)=0$.

On the other hand, the last term of (\ref{WT}) is evaluated by a similar
method for the four-dimensional case by 
Kikukawa and Yamada\cite{weakexpansion},
and we obtain
\begin{equation}
a{\rm{Tr}}[\gamma_5D(n,n)]={i \over \pi\sqrt{N}}\sum_{\mu\nu}\epsilon_{\mu\nu}
\partial_\nu\lambda_\mu.
\label{anomaly} 
\end{equation}
Equation (\ref{anomaly}) is nothing but the chiral anomaly
in two dimensions.

Then the final form of the WT identity is given by
\begin{eqnarray}
\partial_\mu j^k_{5,\mu}&=&i\delta^{k0}{\sqrt{NL} \over \pi}
\sum_{\mu\nu}\epsilon_{\mu\nu}\partial_\nu\lambda_\mu+2M\epsilon\sqrt{N}
\varphi^k_5  \nonumber   \\
&=&i\delta^{k0}{\sqrt{NL} \over \pi}
\sum_{\mu\nu}\epsilon_{\mu\nu}\partial_\nu\lambda_\mu+\sqrt{{2N\over \pi}}
m^2_\pi\times {\varphi^k_5\over \sqrt{2\pi M^2}}.
\label{WT2}
\end{eqnarray}
Then it is obvious that the PCAC relation is satisfied in the
overlap fermion formalism.

In this paper, we studied the overlap fermion formalism by using 
the two-dimensional gauged Gross-Neveu model in the large-$N$ limit,
and showed that the pion mass is automatically proportional to the bare 
quark mass (i.e., the current quark mass) without any fine tuning 
and that the PCAC relation is satisfied.
This result means that the chiral limit of the overlap fermion
formalism is reached by $M_B \rightarrow 0$\cite{Hasenfratz}.
This is in sharp contrast to the Wilson fermion formalism
in which fine tuning of the Wilson parameter is required,
and it is expected that the overlap fermion is quite useful
for numerical studies of lattice QCD$_4$ and other realistic theories.

\bigskip
{\bf Acknowledgments}  \\
KN, one of the authors, would like to thank Prof.T.Yoneya for 
useful discussions.
\newpage

 %
 %
 %
\end{document}